\documentclass[twocolumn]{aastex62}

\turnoffediting

\graphicspath{{./}{figures/}}

\received{2018 July 26}
\revised{2018 November 6}
\accepted{2018 November 14}
\submitjournal{ApJ}

\shorttitle{ALMA, ATCA, and Spitzer Observations of SN 1978K}
\shortauthors{Smith et al.}

\newcommand\ROSAT{{\it ROSAT}}

\newcommand\Spitzer{{\it Spitzer}}

\newcommand\AKARI{{\it AKARI}}

\begin{document}

\title{ALMA, ATCA, and Spitzer Observations of the Luminous
Extragalactic Supernova SN 1978K}

\correspondingauthor{Ian Smith}
\email{iansmith@rice.edu}

\author[0000-0001-8605-5608]{I. A. Smith}
\affiliation{Department of Physics and Astronomy, Rice University, \\
6100 South Main, MS-108, Houston, TX 77251-1892, USA}

\author[0000-0003-4501-8100]{S. D. Ryder}
\affiliation{Australian Astronomical Observatory, 105 Delhi Road, North Ryde, NSW 2113, Australia}
\affiliation{Department of Physics and Astronomy,
Macquarie University, NSW 2109, Australia}

\author[0000-0001-5455-3653]{R. Kotak}
\affiliation{Department of Physics and Astronomy, University of Turku,
Vesilinnantie 5, FI-20014, Finland}

\author[0000-0002-7252-3877]{E. C. Kool}
\affiliation{Department of Physics and Astronomy,
Macquarie University, NSW 2109, Australia}
\affiliation{Australian Astronomical Observatory, 105 Delhi Road, North Ryde, NSW 2113, Australia}

\author{S. K. Randall}
\affiliation{European Southern Observatory, 
Karl-Schwarzschild-Stra{\ss}e 2, D-85748, Garching, Germany}




\begin{abstract}

Only three extragalactic supernovae have been detected at late times 
at millimeter wavelengths: SN~1987A, SN~1978K, and SN~1996cr. 
SN~1978K is a remarkably luminous Type~IIn supernova that 
remains bright at all wavelengths 40 years after its explosion. 
Here we present Atacama Large Millimeter/submillimeter Array (ALMA) 
observations taken in 2016 using Bands 3, 4, 6, and 7 that show a 
steepening in the spectrum.
An absorbed single power law model broadly fits all the radio and
millimeter observations, but would require significant chromatic 
variability.
Alternatively, a broken power law fits the radio-millimeter spectrum:
this can be explained using an ultra-relativistic spherical 
blast wave in a wind scaling with a cooling break, as in a 
gamma-ray burst afterglow.
Using updated Australia Telescope Compact Array (ATCA) light curves, 
we show the non-thermal radio continuum continues to decay as $t^{-1.53}$; 
in the fireball model, this independently defines the power law indices 
found in the radio-millimeter spectrum.
Supernovae such as SN~1978K might be important contributors 
to the Universal dust budget: only SN~1978K was detected in a search 
for warm dust in supernovae in the transitional phase (age 10--100 years). 
Using {\it Spitzer Space Telescope} observations, we show that at
least some of this dust emission has been decaying rapidly as $t^{-2.45}$ 
over the past decade, suggesting it is being destroyed.
Depending on the modeling of the synchrotron emission, the ALMA 
observations suggest there may be emission from a cold 
dust component.

\end{abstract}



\section{Introduction} \label{sec:intro}

\object{SN 1978K} was only the second supernova to be detected and recognized 
as a supernova from its radio emission and the first from its X-rays
\citep{Ryder1993}.
\edit1{Although the date of the explosion ($t_0$) remains unknown, 
we assume it was 1978-05-22 (MJD 43650), as adopted by 
\citet{Montes1997}.}
It is rare that supernovae are bright enough to be followed in 
the X-ray band \citep{Bregman2003}, and SN~1978K is one of only 
a few that have had long-term multi-wavelength monitoring; it 
remains bright at all wavelengths 40 years after the explosion.

SN~1978K lies in the nearby late-type barred spiral galaxy NGC~1313.
This appears to be an isolated galaxy at a distance of only 
$\sim 4.4 - 4.6$~Mpc \citep{Jacobs2009, Qing2015}.
It has undergone vigorous irregular star formation, driven in part by 
the action of numerous \ion{H}{1} supershells \citep{Ryder1995}.
The disk is inclined at 48\arcdeg\ to the line of sight -- permitting 
an excellent view of the whole galaxy -- and the diffuse X-ray emission 
is low; this has allowed detailed studies of its point sources.

As the supernova blast wave expands, we can study (on a human time scale)
how the wind from the massive progenitor evolved in its final hundreds 
and thousands of years.
Besides being historically interesting, SN~1978K has many intriguing 
properties in its own right.  

The peak radio luminosity was very high; at its peak, SN~1978K would 
have been one of the most luminous radio supernovae ever. 
SN~1978K is thus one of the most important members of the ``Type~IIn'' 
sub-class of supernovae, 
\citep[e.g.][]{Schlegel1990,Schlegel2000,Gal-Yam2007,Smith2007b,
Bauer2008,Chandra2012a,Chandra2012b}.
Type~IIn events make up $\sim 10$\% of all core-collapse supernovae 
\citep{Smith2011}.

In \citet{Ryder2016}, we showed a radio VLBI image from 
2015 March 29 at 8.4~GHz.
This revealed that the source remains compact -- with a $< 5$~mas (0.11~pc) 
diameter -- allowing us to place an upper limit on the average expansion
velocity of $1500~{\rm km~s}^{-1}$.
This is consistent with there being a dense circumstellar medium 
surrounding the progenitor star.

The optical spectrum of SN~1978K is dominated by emission lines that 
are only moderately broad, unlike most supernova ejecta 
\citep{Ryder1993,Chugai1995,Chu1999,Kuncarayakti2016}.
This also shows that the surrounding medium was dense, causing a 
rapid slowdown of the shock.
The possibility of a strong wind from the progenitor led
to the suggestion that it was a Luminous Blue Variable 
\citep{Chu1999,Gruendl2002,Gal-Yam2007,Kiewe2012}.
Although no pre-explosion imaging is available that might have allowed 
the progenitor properties to be determined, age-dating of the stellar 
population surrounding SN~1978K using the {\it Hubble Space Telescope}
suggests a progenitor mass of $8.8 \pm 0.2~ M_\sun$
\citep{Williams2018}.

\citet{Kuncarayakti2016} compared SN~1978K spectra taken from 1990 
through 2014 and found that different lines evolved differently.
This suggests the evolution is not spherically symmetric, similar to
SN~1987A.

While the radio flux dropped steadily, the X-ray and UV/optical fluxes 
for SN~1987K remained surprisingly constant from 2000 to at least 2008 
\citep{Smith2007a}.
Starting in 2013, our X-ray, UV, and optical observations 
finally revealed these were fading, but with chromatic variability
\citep{Zhao2017,Smith2019}.
As noted by \citet{Dwarkadas2012}, many SN Type~IIn light curves
do not show the simple decaying X-ray luminosity that would be expected 
from a blast wave expanding into a steady wind.

Over many years, we have been performing 
detailed radio through X-ray observations of SN~1978K
\citep{Smith2007a,Smith2019,Ryder2016}.
In this paper, we focus on new millimeter observations made by the 
Atacama Large Millimeter/submillimeter Array (ALMA), radio 
observations made by the Australia Telescope Compact Array (ATCA),
and infrared observations made by the {\it Spitzer Space Telescope}
(\Spitzer).
In Section \ref{sec:obs}, we present the ALMA millimeter observations 
of SN~1978K from 2016, updated ATCA radio light curves, and \Spitzer\
infrared light curves and a spectrum.
In Section \ref{sec:model}, we use two different models to explain the
combined radio-millimeter synchrotron spectrum.
In Section \ref{sec:dust}, we discuss the implications of the new
observations on the warm and cold dust emission in SN~1978K.
In Section \ref{sec:conclusion}, we outline how future observations
might reveal more information about the source.

\section{Observations and Results} \label{sec:obs}

\subsection{2016 ALMA Observations of SN~1978K} \label{subsec:alma}

Only three extragalactic supernovae have been detected at late times 
at millimeter wavelengths: SN~1978K, 
SN~1987A \citep{Lakicevic2012,Matsuura2011,Matsuura2015},
and SN~1996cr \citep{Meunier2013}.
SN~1978K was first detected at 34 and 94~GHz in our ATCA observations 
in 2014 September \citep{Ryder2016}.
SN~1987A is only detectable because it is nearby; 
at the distance of SN~1978K, it would not be detectable by ALMA.
This highlights the danger of extrapolating the SN~1987A results to 
other supernovae, and shows the importance of studying other luminous 
supernovae.

During Cycle 3 in 2016, ALMA observed SN~1978K in Bands 3, 4, 6, and 7.
Details of the observations are given in Table \ref{table:almadates}.
SN~1978K was observed at a good elevation in all four bands.
The weather conditions were stable during all the observations.
The sources J0519--4546 (Bands 3, 4, and 6), J2258--2758 (Band 7), and 
J0538--4405 (Band 7) were used as bandpass calibrators; 
J0334--4008 (Band 3), J0519-4546 (Bands 4 and 6), and Ceres (Band 7)
were used as amplitude calibrators; 
and J0303--6211 was the phase calibrator for all the observations.

\begin{deluxetable*}{cccccccc}
\tablecaption{
ALMA millimeter observations of SN~1978K in 2016
\label{table:almadates}
}
\tablehead{
\colhead{Band} &
\colhead{UT Date} &
\colhead{MJD} &
\colhead{Age\tablenotemark{a}} & 
\colhead{Time on} &
\colhead{Maximum} &
\colhead{Precipitable} &
\colhead{Average Source} \\ 
\colhead{Number} &
\colhead{(YYYY-MM-DD)} &
\colhead{} &
\colhead{(days)} &
\colhead{Science Target} &
\colhead{Baseline} &
\colhead{Water Vapor} &
\colhead{Elevation} \\
\colhead{} &
\colhead{} &
\colhead{} &
\colhead{} &
\colhead{(min)} &
\colhead{(km)} &
\colhead{(mm)} &
\colhead{(degrees)} 
}
\startdata
 3 & 2016-09-02 & 57633 & 13983 & 16.7 & 1.8 & 1.38 & 42.4 \\
 4 & 2016-07-25 & 57594 & 13944 & 21.2 & 1.1 & 0.94 & 45.9 \\
 6 & 2016-06-09 & 57549 & 13899 & 26.8 & 0.7 & 1.23 & 45.5 \\
 7 & 2016-06-16 & 57556 & 13906 & 70.4 & 0.8 & 0.72 & 44.1 \\
\enddata
\tablenotetext{a}{Age based on the adopted explosion date of 1978-05-22
(MJD 43650).}
\end{deluxetable*}

All the observations were reprocessed using the complete calibration
pipeline in CASA version 4.7.2 \citep{casa}.
Multi-frequency synthesis with the mtmfs deconvolver was used when 
combining channels to make continuum images; $\rm{nterms}=2$ was used,
which assumes a simple sloped spectrum.
The images were cleaned using the task TCLEAN with a Briggs weighting
and {\tt robust=0.5}.
Primary beam corrections were applied on the restored images.
The observations were short, and the source was not bright enough to 
merit doing phase or amplitude self calibration.

SN~1978K was well detected in the observations in all four bands.
The background around the source was smooth.
SN~1978K was consistent with being point-like in all the observations.

\begin{figure}[ht!]
\plotone{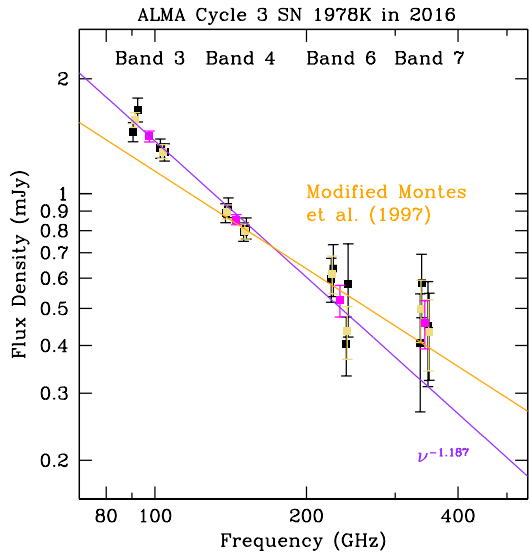}
\caption{The Cycle 3 ALMA observations of SN~1978K in 2016.
The black points show the four separate spectral windows for each band.
The gold points combine the spectral windows in pairs, and the
magenta points combine all the data for each band.
The orange curve shows the modified version of the \citet{Montes1997} 
model described in Section \ref{subsec:montes}.
The purple curve shows the $\nu^{-1.187}$ power law appropriate
for an ultra-relativistic spherical blast wave with a wind scaling
above the cooling break that is described in Section \ref{subsec:fireball}.
}
\label{figure:alma}
\end{figure}

Each ALMA band is split into four spectral windows (SPWs) of bandwidth 
1.875~GHz.
Each SPW contains 128 channels.
No obvious lines were found in addition to the continuum.
Thus all the good channels were combined to obtain the continuum
fluxes in the four separate SPW.
These results are shown in Table \ref{table:almafluxes} and as the
black points in Figure \ref{figure:alma}.

\begin{deluxetable*}{ccCCCc}
\tablecaption{
ALMA results for SN~1978K in 2016
\label{table:almafluxes}
}
\tablehead{
\colhead{Band} &
\colhead{Spectral} &
\colhead{Center Frequency} &
\colhead{Beam} &
\colhead{Source Flux Density} &
\colhead{Image RMS} \\ 
\colhead{Number} &
\colhead{Window(s)} &
\colhead{of Image} &
\colhead{(arcsec)} &
\colhead{(mJy)} &
\colhead{(mJy/beam)} \\
\colhead{} &
\colhead{} &
\colhead{(GHz) (LSRK)} &
\colhead{} &
\colhead{} &
\colhead{} 
}
\decimals
\startdata
 3 & 17      & \phn90.4832 & 0.66 \times 0.46 & 1.451 \pm 0.083 & 0.048 \\
   & 19      & \phn92.4286 & 0.64 \times 0.47 & 1.660 \pm 0.120 & 0.065 \\
   & 21      &    102.5150 & 0.58 \times 0.40 & 1.314 \pm 0.077 & 0.045 \\
   & 23      &    104.5073 & 0.57 \times 0.40 & 1.284 \pm 0.067 & 0.044 \\
   & 17 + 19 & \phn91.2567 & 0.65 \times 0.46 & 1.561 \pm 0.070 & 0.041 \\
   & 21 + 23 &    103.5151 & 0.57 \times 0.40 & 1.289 \pm 0.054 & 0.031 \\
   & All 4   & \phn97.4992 & 0.59 \times 0.42 & 1.412 \pm 0.047 & 0.025 \\
\hline
 4 & 17      &    138.1083 & 0.57 \times 0.41 & 0.890 \pm 0.051 & 0.032 \\
   & 19      &    139.9364 & 0.56 \times 0.43 & 0.920 \pm 0.055 & 0.029 \\
   & 21      &    150.0149 & 0.52 \times 0.37 & 0.794 \pm 0.044 & 0.028 \\
   & 23      &    151.9056 & 0.52 \times 0.37 & 0.812 \pm 0.052 & 0.031 \\
   & 17 + 19 &    138.9833 & 0.55 \times 0.41 & 0.896 \pm 0.038 & 0.022 \\
   & 21 + 23 &    150.9055 & 0.51 \times 0.37 & 0.796 \pm 0.036 & 0.021 \\
   & All 4   &    145.0147 & 0.52 \times 0.39 & 0.855 \pm 0.026 & 0.015 \\
\hline
 6 & 17      &    223.9891 & 0.69 \times 0.52 & 0.598 \pm 0.079 & 0.050 \\
   & 19      &    225.9892 & 0.68 \times 0.51 & 0.636 \pm 0.099 & 0.046 \\
   & 21      &    240.0207 & 0.63 \times 0.47 & 0.404 \pm 0.071 & 0.052 \\
   & 23      &    242.0207 & 0.63 \times 0.46 & 0.580 \pm 0.159 & 0.059 \\
   & 17 + 19 &    224.9891 & 0.67 \times 0.51 & 0.616 \pm 0.070 & 0.035 \\
   & 21 + 23 &    241.0207 & 0.62 \times 0.46 & 0.437 \pm 0.069 & 0.040 \\
   & All 4   &    233.0049 & 0.63 \times 0.48 & 0.525 \pm 0.050 & 0.026 \\
\hline
 7 & 0       &    336.5019 & 0.37 \times 0.30 & 0.407 \pm 0.139 & 0.060 \\
   & 1       &    338.4394 & 0.37 \times 0.30 & 0.583 \pm 0.110 & 0.053 \\
   & 2       &    348.5021 & 0.36 \times 0.29 & 0.450 \pm 0.138 & 0.056 \\
   & 3       &    350.5022 & 0.36 \times 0.29 & 0.436 \pm 0.112 & 0.065 \\
   & 0 + 1   &    337.4706 & 0.37 \times 0.30 & 0.500 \pm 0.093 & 0.042 \\
   & 2 + 3   &    349.5022 & 0.36 \times 0.29 & 0.435 \pm 0.092 & 0.043 \\
   & All 4   &    343.5020 & 0.36 \times 0.29 & 0.458 \pm 0.066 & 0.030 \\
\enddata
\end{deluxetable*}

The gold points in Figure \ref{figure:alma} combine the SPW in pairs;
there is only a small difference in the frequencies for the two lower
SPW and for the two higher SPW, making these combinations robust.

The magenta points in Figure \ref{figure:alma} combine all the SPW in 
each band.
There is a larger difference in the frequencies for the lower and
higher pairs of SPW.
Given changes in the beam across the broad bands and interpolation 
concerns, it might be considered a little less robust to combine all 
the data for the whole band.
However, the magenta points show this can be done without any problems.

\subsection{Updated ATCA Light Curves for SN~1978K} \label{subsec:atca}

We have been performing radio monitoring of SN~1978K using the Australia 
Telescope Compact Array (ATCA) at irregular intervals since its discovery
\citep{Ryder1993,Schlegel1999,Smith2007a}.

The Compact Array Broad-band Backend \citep[CABB;][]{Wilson2011}
was commissioned in early 2009.
It provides $2 \times 2$~GHz IF bands, each of which has $2 \times 2048$
channels of 1~MHz each.
This yields a factor of 4 improvement in the continuum
sensitivity over the 128~MHz bandwidths used for the observations reported
in our previous papers.
The band central frequencies are chosen to minimize radio frequency
interference across each band, and differ from those shown in
our previous work. 
Details of the ATCA CABB observations are given in Table \ref{table:atca}.
We remark that some pre-CABB ATCA observations of SN~1978K taken 
between 2007-11-21 and 2009-01-29 suffered from problems with the
phase stability and are not included here.

\begin{deluxetable*}{ccccCCCCCC}
\tablecaption{
ATCA CABB radio observations of SN~1978K
\label{table:atca}
}
\tablehead{
\colhead{UT Date} &
\colhead{MJD} &
\colhead{Age\tablenotemark{a}} & 
\colhead{Array} & 
\multicolumn{6}{c}{Flux Densities (mJy)} \\
\colhead{(YYYY-MM-DD)} & 
\colhead{} &
\colhead{(days)} &
\colhead{} &
\colhead{1.33 GHz} & 
\colhead{1.84 GHz} & 
\colhead{2.36 GHz} & 
\colhead{2.87 GHz} & 
\colhead{5.50 GHz} & 
\colhead{9.00 GHz}
}
\startdata
2009-10-09 & 55114 & 11464 & H75C  & \nodata    & \nodata    & \nodata    & \nodata    & 19.2\pm2.2 & 10.5\pm3.3 \\
2009-10-13 & 55118 & 11468 & H168C & \nodata    & \nodata    & 37.6\pm6.2 & \nodata    & \nodata    & \nodata    \\
2013-06-07 & 56451 & 12801 & 6C    & 31.4\pm2.7 & 26.1\pm1.7 & 23.0\pm1.3 & 20.7\pm0.9 & 15.1\pm0.8 & 10.7\pm0.5 \\
2017-09-04 & 58001 & 14351 & 1.5A  & \nodata    & \nodata    & \nodata    & \nodata    & 12.1\pm0.5 &  8.4\pm0.5 \\
2018-03-19 & 58197 & 14547 & EW352 & 35.7\pm4.7 & 27.6\pm1.4 & 22.2\pm1.3 & 18.6\pm2.5 & 12.3\pm0.6 &  8.5\pm1.0 \\
\enddata
\tablenotetext{a}{Age based on the adopted explosion date of 1978-05-22 
(MJD 43650).}
\end{deluxetable*}

The ATCA primary flux calibrator, PKS~B1934--638, was observed once
per run at each frequency to set the absolute flux scale, as well as
define the bandpass calibration in each band. 
Frequent observations of the nearby source PKS~0302--623 allowed us to 
monitor and correct for variations in gain and phase throughout each run. 

\begin{figure}[ht!]
\plotone{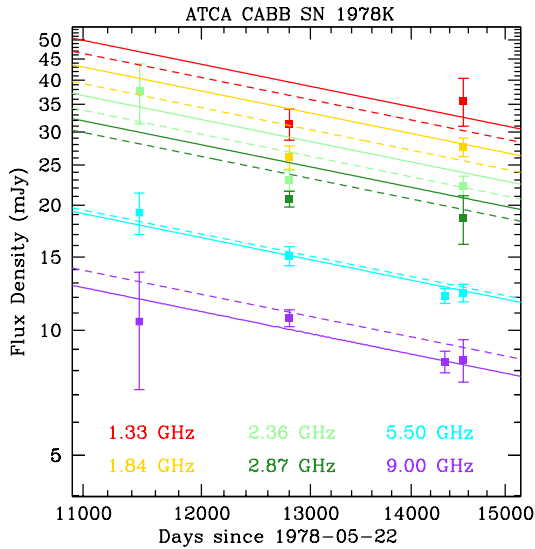}
\caption{The radio light curves for SN~1978K from Australia Telescope 
Compact Array monitoring using the Compact Array Broad-band Backend 
between 2009 and 2018.
From top to bottom, these are at 1.33~GHz (red), 1.84~GHz (gold),
2.36~GHz (light green), 2.87~GHz (dark green), 5.50~GHz (cyan),
and 9.00~GHz (purple).
The solid curves show the evolution given by the modified version of 
the \citet{Montes1997} model described in Section \ref{subsec:montes};
this asymptotes to a $t^{-1.53}$ power law at late times for all 
the frequencies.
The dashed curves use the absorbed ultra-relativistic spherical blast 
wave with a wind scaling model described in Section \ref{subsec:fireball}:
$A = 90.3$, $B = 3.287 \times 10^5$, and $K_4 = 7.0 \times 10^{-3}$.
1978-05-22 was MJD 43650.}
\label{figure:atca}
\end{figure}

The data were processed using the Miriad package \citep{miriad} 
and using procedures outlined in Sec.~4.3 of the ATCA User 
Guide\footnote{http://www.narrabri.atnf.csiro.au/observing/users\_guide/html/atug.html}. 
After editing and calibrating the data, images at each frequency were 
made using robust weighting ({\tt robust=0.5}), then cleaned
down to 3$\times$ the r.m.s. noise level.
The ``16 cm'' band covers 1.1--3.1~GHz, or a factor of 3 in beam size, 
so has to be divided into 4 sub-bands of 512~MHz each in order to be 
processed properly.
Fitting a Gaussian point source at the location of SN~1978K
yielded the flux densities shown in Table \ref{table:atca} and 
Figure \ref{figure:atca}.
The uncertainty in each case is a combination of the fitting
and absolute flux calibration errors. 

\subsection{\Spitzer\ Observations of SN~1978K} \label{subsec:spitzer}

\Spitzer\ observations of younger supernovae (age $< 10$~years)
have shown that Type~IIn can initially be bright in the mid-infrared
\citep[e.g.][]{Szalai2018}.

\citet{Tanaka2012} used \Spitzer\ and \AKARI\ to study 6 supernovae 
in the ``transitional'' phase (age $10-100$~years).
Only SN~1978K was detected, in 2006.
This is suggestive of emission from warm (a few 100~K) dust.  

Since 2006, the InfraRed Array Camera (IRAC) on \Spitzer\ 
\citep{Fazio2004,Werner2004} has continued to make observations of 
NGC~1313 in both the cryogenic and post-cryogenic phases; 
a few of these observations have had SN~1978K in the field of view.
Details of the observations are given in Table \ref{table:spitzer}.

\begin{deluxetable*}{ccccCCCC}
\tablecaption{
\Spitzer\ IRAC infrared observations of SN~1978K
\label{table:spitzer}
}
\tablehead{
\colhead{UT Date} &
\colhead{MJD} &
\colhead{Age\tablenotemark{a}} &
\colhead{AOR} &
\multicolumn{4}{c}{Flux Densities (mJy)} \\
\colhead{(YYYY-MM-DD)} &
\colhead{} &
\colhead{(days)} &
\colhead{} &
\colhead{3.6 \micron} &
\colhead{4.5 \micron} &
\colhead{5.8 \micron} &
\colhead{8.0 \micron}
}
\startdata
2007-09-12 & 54356 & 10706 & 22524160\tablenotemark{b} &
0.05764\pm0.00124 & 0.10620\pm0.00130 & 0.1831\pm0.0065 & 0.7347\pm0.0072 \\
2008-06-10 & 54627 & 10977 & 26537216 &
0.05353\pm0.00180 & 0.09883\pm0.00262 & 0.1796\pm0.0103 & 0.6971\pm0.0129 \\
2011-07-07 & 55749 & 12099 & 42193408 &
0.04475\pm0.00188 & 0.07783\pm0.00132 & \nodata         & \nodata \\
2012-02-01 & 55959 & 12309 & 42193152 &
0.04169\pm0.00195 & 0.07623\pm0.00126 & \nodata         & \nodata \\
2014-02-12 & 56701 & 13051 & 50504704 &
\nodata           & 0.07168\pm0.00205 & \nodata         & \nodata \\
2016-08-05 & 57606 & 13956 & 52691456 &
0.03138\pm0.00083 & \nodata           & \nodata         & \nodata \\
2017-01-02 & 57755 & 14105 & 60812544 &
\nodata           & 0.05311\pm0.00063 & \nodata         & \nodata \\
2018-02-09 & 58158 & 14508 & 60813568 &
\nodata           & 0.05132\pm0.00065 & \nodata         & \nodata \\
2018-08-24 & 58355 & 14705 & 66130688 & 
0.02500\pm0.00068 & 0.04858\pm0.00058 & \nodata         & \nodata \\
\enddata
\tablenotetext{a}{Age based on the adopted explosion date of 1978-05-22 
(MJD 43650).}
\tablenotetext{b}{Combines AOR 22524160 and 22524416.}
\end{deluxetable*}

\edit1{The IRAC observations were processed using the most recent
\Spitzer\ pipelines; these were versions S18.25 for the cryogenic 
observations (prior to 2009 May) and S19.2 for the post-cryogenic 
observations.}
SN~1978K was well detected in all four channels.
The surrounding background is fairly smooth and free from other sources.
Point source photometry was performed on the Corrected Basic Calibrated 
Data (CBCD) generated by the pipeline using APEX in the MOPEX package
\citep{mopex}.
Short exposure frames were dropped from the mosaicing.
Location-dependent photometric corrections that are often applied to 
bluer stars were not used for SN~1978K, since it is a very red source.
A Point Response Function (PRF) map was used in APEX Multiframe to 
improve the fitting for sources outside the central region.
Correction factors for the PRF flux densities were taken from Table C1
of the IRAC Instrument Handbook version 2.1.2.

\begin{figure}[ht!]
\plotone{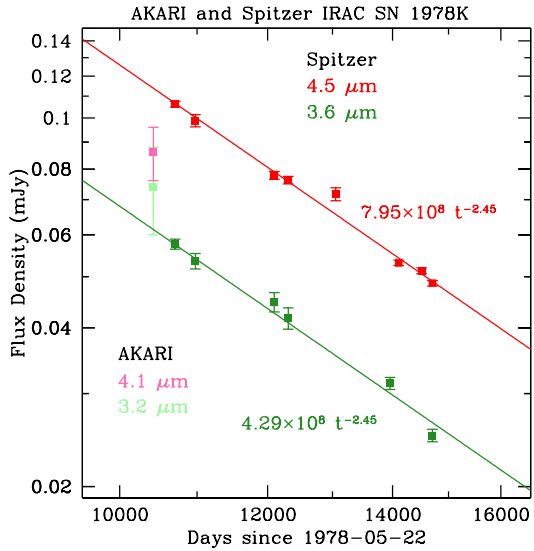}
\caption{The infrared light curves for SN~1978K 
at 4.5~\micron\ (red, upper) and 3.6~\micron\ (dark green, lower) 
from \Spitzer\ IRAC observations from 2007 to 2018.
A $t^{-2.45}$ power law decay is shown for both channels.
The 2006 December 03--04 \AKARI\ observations \citep{Tanaka2012} are 
included for comparison, although these use different wavelength bands: 
4.1~\micron\ (pink, upper) and 3.2~\micron\ (light green, lower).
1978-05-22 was MJD 43650.}
\label{figure:irac}
\end{figure}

The \Spitzer\ IRAC light curves for SN~1978K are shown in 
Figure \ref{figure:irac} and reveal a rapid fading of the source.
The fluxes of other nearby stars of similar brightness were consistent 
with remaining constant over the decade (to within the expected few 
percent uncertainties).

On 2008-07-07 (MJD 54654, AOR 26536704), the InfraRed Spectrograph 
(IRS) on \Spitzer\ \citep{Houck2004} made cryogenic observations 
pointed at SN~1978K using the Staring Mode in Channel 0 (Short-Low) 
and Channel 2 (Long-Low).
The IRS Enhanced Spectrophotometric Products were produced starting 
with the cryogenic pipeline S18.18 processed spectra and 
merged using IRS MERGE v2.1 to produce the spectrum
SPITZER\_S5\_26536704\_01\_merge.tbl in the \Spitzer\ Heritage Archive.
Data that were flagged as questionable were removed.

\begin{figure}[ht!]
\plotone{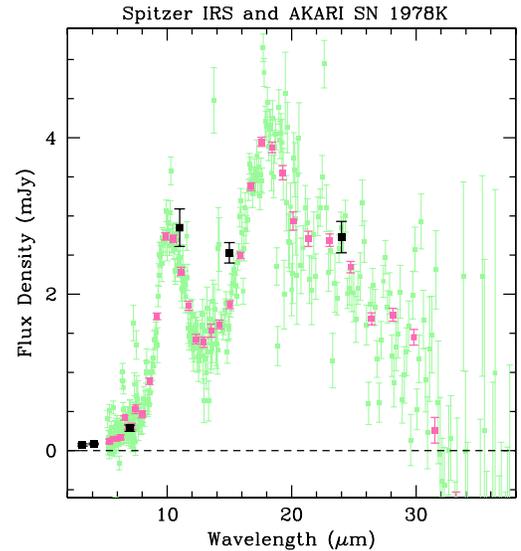}
\caption{The mid-infrared spectrum of SN~1978K taken 2008-07-07
by the \Spitzer\ IRS using the full spectral resolution (light green)
and re-binned using weighted means to combine every ten data points (pink).
The 2006 December 03--04 \AKARI\ observations \citep{Tanaka2012} are 
included for comparison (black).}
\label{figure:irs}
\end{figure}

The \Spitzer\ IRS mid-infrared spectrum is shown in Figure \ref{figure:irs};
see also \citet{vandyk2011}.
The double-peaked spectrum is broadly consistent with the \AKARI\ 
observations and modeling shown in \citet{Tanaka2012}.

\section{Radio-Millimeter Spectral Modeling} \label{sec:model}

\begin{figure}[ht!]
\plotone{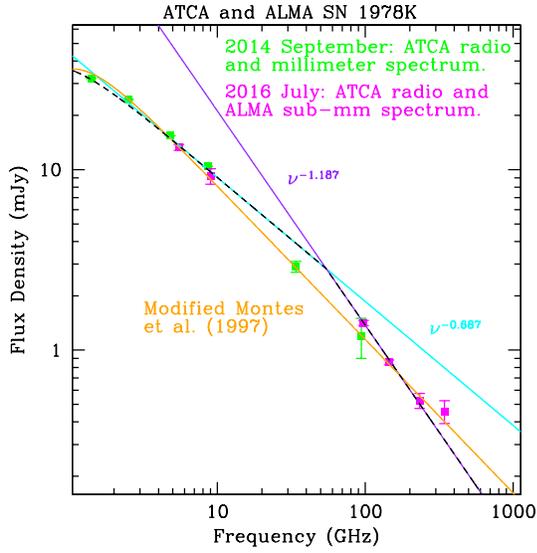}
\caption{The green points reproduce Figure 2 of \citet{Ryder2016} 
showing our ATCA millimeter observations from 2014 September along 
with the radio spectrum extrapolated to that time.
The magenta points show our 2016 ALMA observations for the full
bands along with ATCA radio points interpolated to that time.
The orange curve shows the modified version of the \citet{Montes1997} 
model described in Section \ref{subsec:montes}.
The power laws show an unabsorbed ultra-relativistic 
spherical blast wave with a wind scaling below the cooling break 
($\nu^{-0.687}$; cyan), and above the cooling break 
($\nu^{-1.187}$; purple) as described in Section \ref{subsec:fireball}.
The black dashed curve uses the absorbed ultra-relativistic spherical blast 
wave with a wind scaling model described in Section \ref{subsec:fireball}:
$A = 90.3$, $B = 3.287 \times 10^5$, and $K_4 = 7.0 \times 10^{-3}$.}
\label{figure:atcaalma}
\end{figure}

Figure \ref{figure:atcaalma} combines the radio and millimeter
observations of SN~1978K, and shows model fits from two 
possible interpretations for the radio-millimeter spectrum.

\subsection{\citet{Montes1997} Model} \label{subsec:montes}

Based on the model of \citet{Weiler1986,Weiler1990} and the early 
observations of SN~1978K, \citet{Montes1997} generated a parameterized 
model for a supernova shock interacting with a high-density ionized 
circumstellar envelope.
\edit1{The model is described in detail in \citet{Montes1997} and
\citet{Schlegel1999}.
This effectively has a single power law spectral index $\alpha$ at 
high frequencies with attenuation at lower frequencies from the 
intervening medium.
The main temporal decay power law index $\beta$ is independent of 
the frequency; $\beta = -1.53$ as shown in the radio decay in 
Figure \ref{figure:atca}.
The initial unabsorbed flux density normalization at 5~GHz is given 
by $K_1$.
There is absorption from local uniform and local non-uniform material
with initial optical depths at 5~GHz given by $K_2$ and $K_3$ and
power law temporal decay indices $\delta \equiv \alpha - \beta -3$ 
and $\delta^{\prime} \equiv 5 \delta / 3$ respectively.
There is also a ``distant'' absorption that is assumed to be
time-independent whose optical depth at 5~GHz is given by $K_4$.
All the absorbing media are assumed to be purely thermal ionized 
hydrogen with the opacities having a $\nu^{-2.1}$ dependence.}

In Figure 2 of \citet{Ryder2016} we modeled the 2014 ATCA radio
and millimeter observations out to 94~GHz using a modified version
of the \citet{Montes1997} model.
The extension to higher frequencies with the ALMA data necessitated
additional tweaking of the model parameters.
The full set of parameters used is given in Table \ref{table:montes}.

\begin{deluxetable}{LC}
\tablecaption{
\citet{Montes1997} model parameters used for SN~1978K in 2016
\label{table:montes}
}
\tablehead{
\colhead{Parameter} &
\colhead{Value}
}
\startdata
\alpha        & -0.85                       \\
\beta         & -1.53                       \\
\delta        & -2.32                       \\
\delta^\prime & -3.87                       \\
t-t_0         & 13934                       \\
K_1           & 3.2 \times 10^7~{\rm mJy}   \\
K_2           & 4.0 \times 10^4             \\
K_3           & 3.6 \times 10^{11}          \\
K_4           & 1.6 \times 10^{-2}          \\
\enddata
\end{deluxetable}

\begin{figure*}[ht!]
\includegraphics[width=0.33\textwidth]{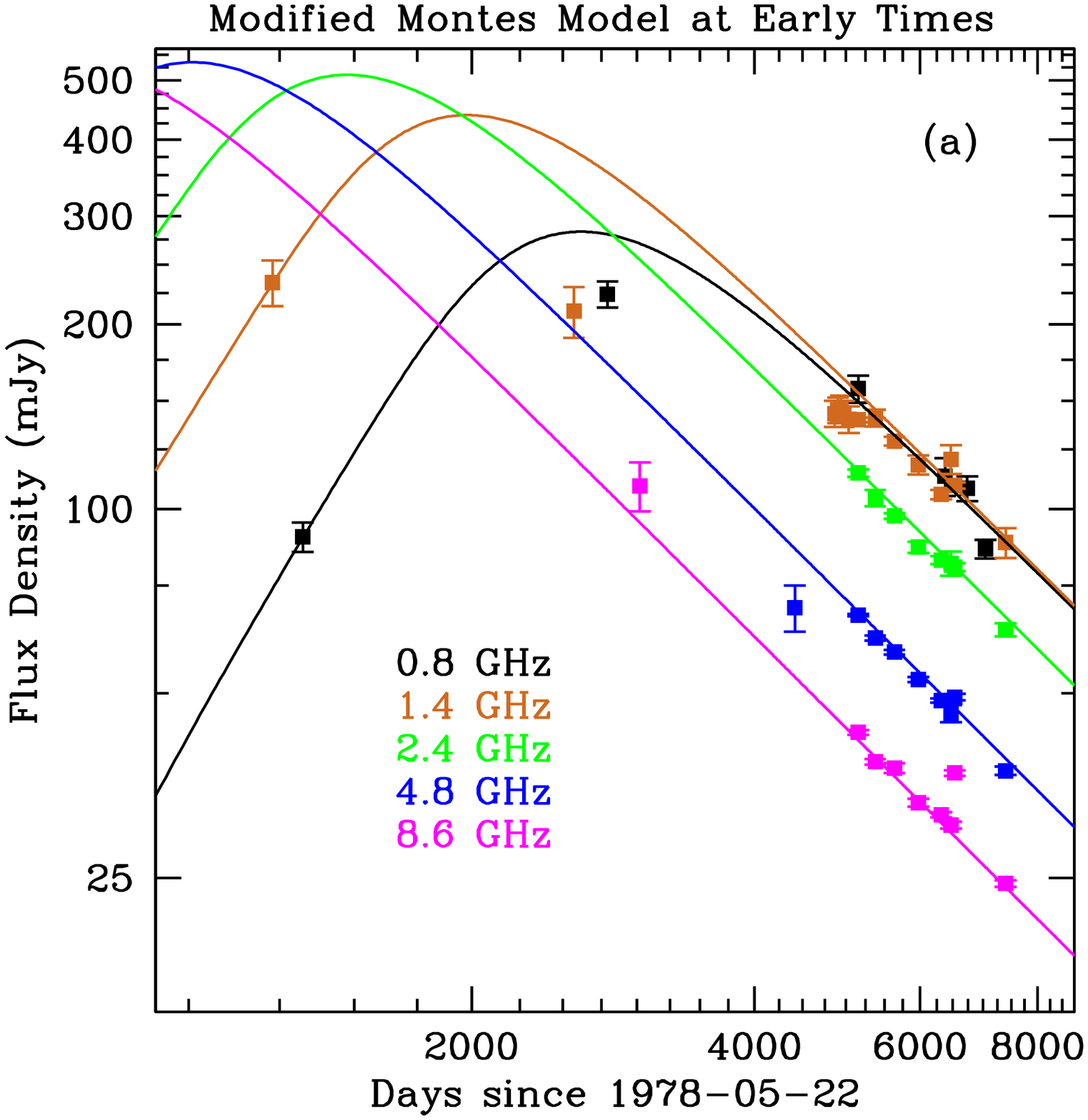}
\includegraphics[width=0.33\textwidth]{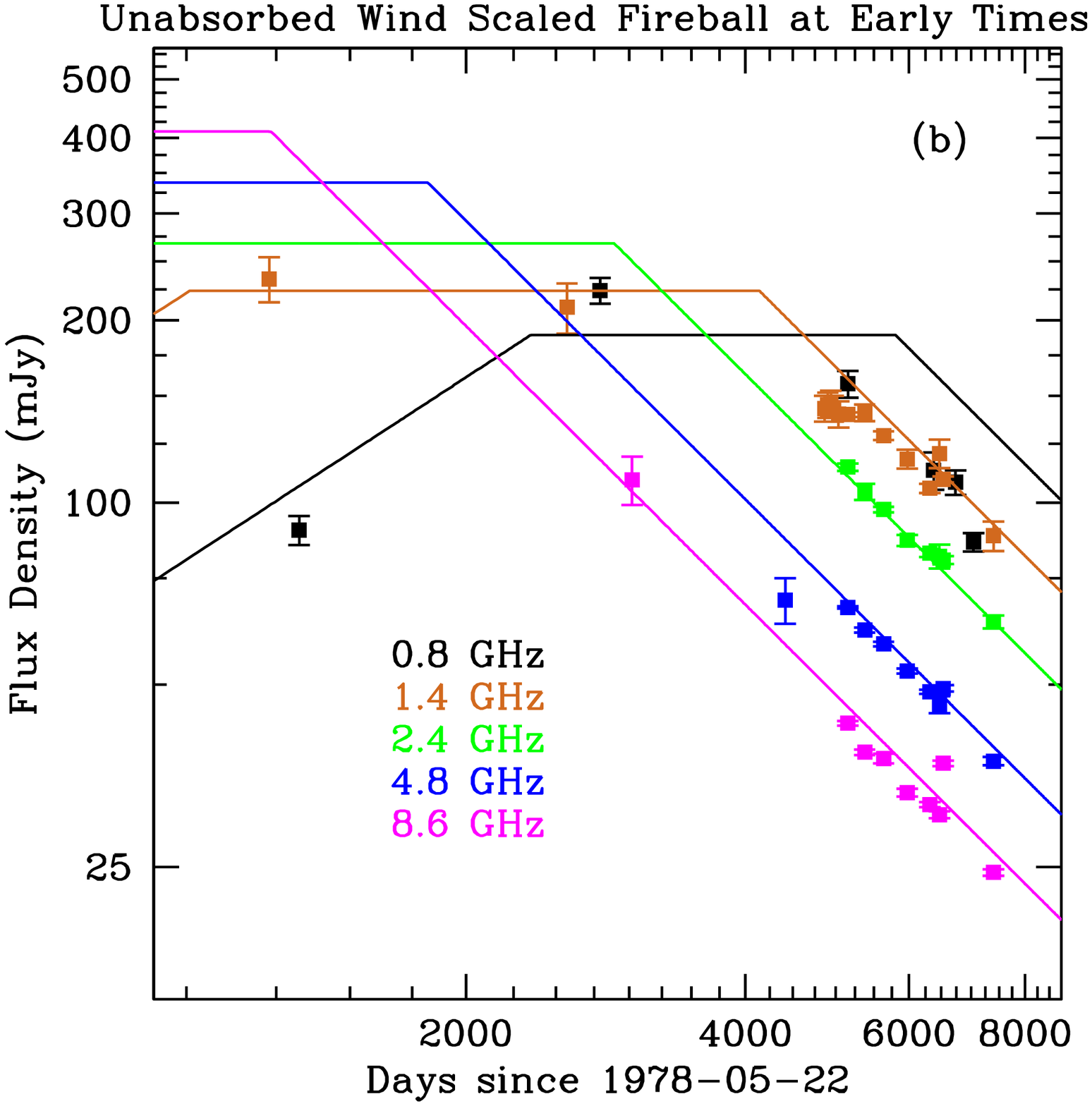}
\includegraphics[width=0.33\textwidth]{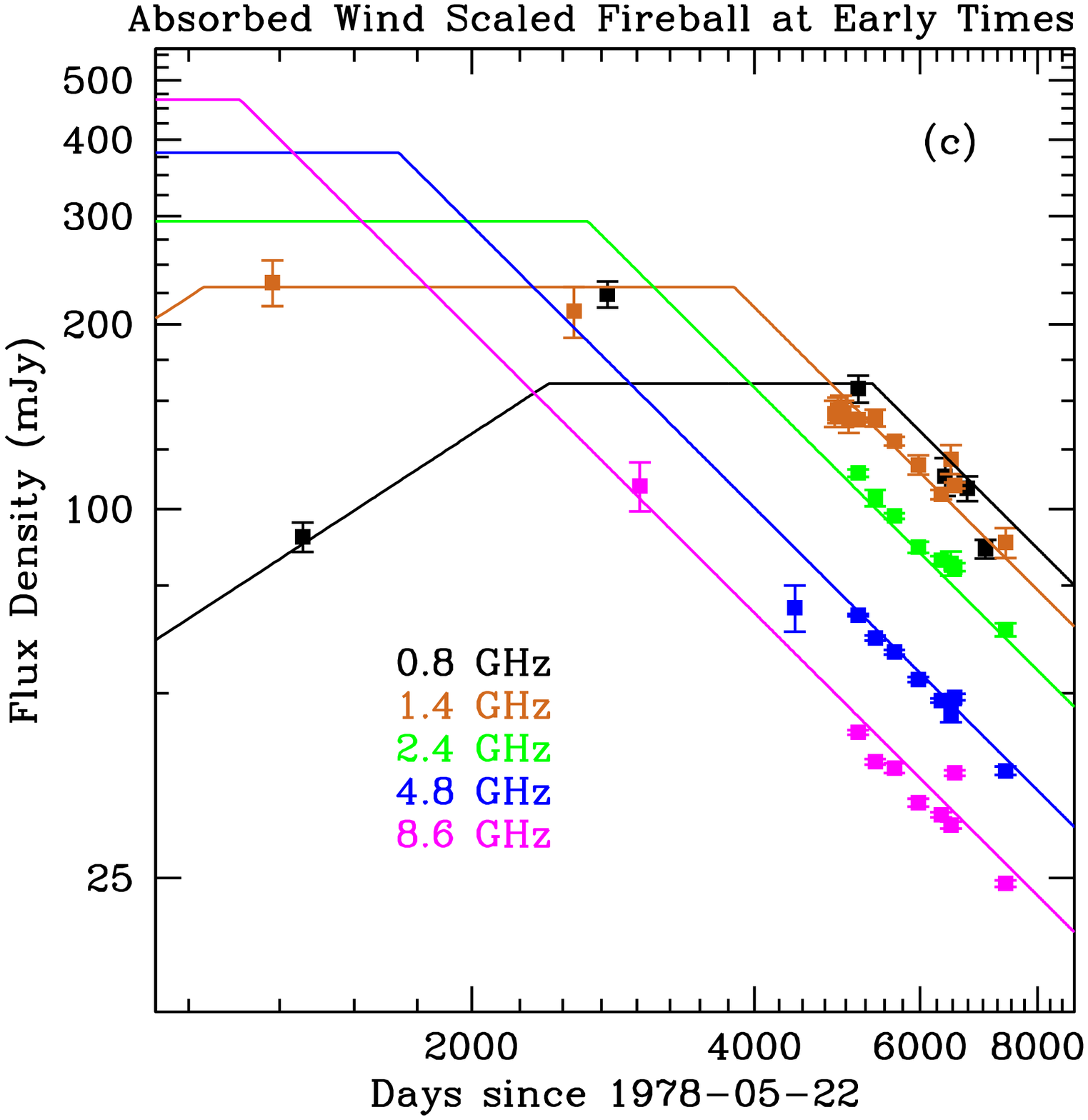}
\caption{The early radio light curves for SN~1978K at 0.8~GHz (black), 
1.4~GHz (brown), 2.4~GHz (green), 4.8~GHz (blue), and 8.6~GHz (magenta).
The data are the same as shown in Figure 10 of \citet{Schlegel1999},
which includes additional observations from \citet{Ryder1993},
\citet{Peters1994}, and \citet{Montes1997}.
(a)
The curves use the modified version of the \citet{Montes1997}
model described in Section \ref{subsec:montes}.
(b)
The curves use the unabsorbed ultra-relativistic spherical blast 
wave with a wind scaling model described in Section \ref{subsec:fireball}:
$A = 88.8$, $B = 3.727 \times 10^5$, and $K_4 = 0$.
(c)
The curves use the absorbed ultra-relativistic spherical blast 
wave with a wind scaling model described in Section \ref{subsec:fireball}:
$A = 90.3$, $B = 3.287 \times 10^5$, and $K_4 = 7.0 \times 10^{-3}$.}
\label{figure:earlyradio}
\end{figure*}

\edit1{As shown in Figure \ref{figure:earlyradio}{(a)}, at early times the
updated \citet{Montes1997} model gives a radio evolution that is
very similar to that shown in Figure 10 of \citet{Schlegel1999}.
As noted in \citet{Schlegel1999}, this model does not have a good 
explanation for the flat initial light curve at 1.4~GHz.}

\edit1{The updated \citet{Montes1997} model is compared to the more recent 
ATCA and ALMA observations described in Section \ref{sec:obs}
using the solid orange curve in Figures \ref{figure:alma} and 
\ref{figure:atcaalma} and the solid colored curves in 
Figure \ref{figure:atca}.
It provides a decent general description of all the recent radio and 
millimeter observations of SN~1978K.}

\edit1{However, the 2013-06-07 ATCA observation appears to have a dip at
the four lower frequencies, but not at the two higher ones;
the data points lie $3.0\sigma$, $4.7\sigma$, $4.7\sigma$, 
and $5.1\sigma$ below the \citet{Montes1997} model curves at 
1.33~GHz, 1.84~GHz, 2.36~GHz, and 2.87~GHz respectively.
The deviation at the lower frequencies went away by 2018-03-19.}

\edit1{Chromatic fluctuations might be the result of inhomogeneities in 
the radio emitting region, implying there were fluctuations in the 
mass-loss rate of the progenitor star as recently as a few hundred 
years before the explosion.}

\edit1{Compact extragalactic radio sources can also appear to vary as a 
result of interstellar scintillation \citep{Walker98}.
SN~1978K is at a galactic latitude of $-44.7\arcdeg$.
For this latitude, the transition frequency at which the scattering
strength is unity $\nu_0 \sim 4.5$~GHz.
Thus this lies in the middle of our range of ATCA radio frequencies.
At this galactic latitude, the angular size limit of the
first Fresnel zone for an extragalactic source being observed at
$\nu_0$ is $\theta_{\rm F0} \sim 5 \times 10^{-6}$~arcsec.
Sources (or components thereof) that are smaller than this size limit
can be approximated as point sources and may show deep modulations
in their received flux.
Our VLBI imaging at 8.4~GHz shows that SN~1978K is $< 5$~mas in diameter.
Thus scintillation could be responsible for a larger variability at the 
lower frequencies if SN~1978K extends over microarcsec scales rather 
than milliarcsec scales.}

\edit1{Figure \ref{figure:alma} shows that the four independent 
Band 3 ALMA SPW all lie well above the \citet{Montes1997} model curve.
The combined data for the whole of Band 3 lies $5.2 \sigma$
above the model.
Thus the \citet{Montes1997} model would need significant chromatic 
fluctuations in the millimeter to explain the excess in 2016.}

\subsection{Broken Power Law Fireball Model} \label{subsec:fireball}

Figure \ref{figure:alma} indicates that the millimeter spectrum
on its own might be better explained using a steeper power law
index than the one used for the radio.
Figure \ref{figure:atcaalma} shows that for a broken power law model, 
the change in the spectral index appears to be $\sim 0.5$.

In a gamma-ray burst (GRB) afterglow spectrum, a change of 0.5 in
the spectral index is expected for a cooling break in the late-time 
emission from an ultra-relativistic spherical blast wave
\citep[e.g.][]{Granot2014}.
In \citet{Smith2007a} we showed that SN~1978K was inside the 
$\sim 4 \sigma$ error box of GRB~771029.
The quality of the GRB locations at that time was poor, and 
this may just have been a chance alignment.
Type~IIn supernovae are not generally expected to produce GRBs, and 
this would have been a very under-luminous burst at the distance of 
NGC~1313.
However, we can still adopt the generic fireball formalism here.

In the GRB fireball model with a wind scaling for the external medium, 
frequencies below the cooling break ($\nu_{\rm c}$) will have a temporal 
decay of $t^{(1-3p)/4}$, where $p$ is the electron spectral index.
Using our observed radio decay of $t^{-1.53}$ from 
Figure \ref{figure:atca}, this implies $p=2.373$.

The fireball model then requires that the spectral indices be
$\nu^{(1-p)/2} = \nu^{-0.687}$ below $\nu_{\rm c}$ and
$\nu^{-p/2} = \nu^{-1.187}$ above it.  
As shown by the cyan and purple curves in Figures \ref{figure:alma} 
and \ref{figure:atcaalma}, these spectral indices naturally explain the 
observed radio-millimeter spectrum.
The fact that both the radio temporal decay and the spectral indices 
independently agree with the fireball expectations make this a promising 
interpretation.

The $\nu^{-0.687}$ and $\nu^{-1.187}$ power laws are the asymptotic 
spectral indices.
In theory, the spectrum could have a sharp break if there is a unique 
cooling frequency. 
In practice, it is usually assumed that there is a smooth transition
between the two power laws.
This would then explain why the 34~GHz ATCA point lies below the cyan
curve in Figure \ref{figure:atcaalma}.

A GRB fireball model with an ISM scaling for the external medium
does not explain the SN~1978K observations.
In this case, the frequencies below $\nu_{\rm c}$ would have a temporal 
decay of $t^{3(1-p)/4} = t^{-1.53}$, giving $p=3.04$.
The spectral indices would then be
$\nu^{(1-p)/2} = \nu^{-1.02}$ below $\nu_{\rm c}$ and
$\nu^{-p/2} = \nu^{-1.52}$ above it: these are both steeper than
in our radio and millimeter observations.

\edit1{Two additional breaks in the spectrum are involved in 
explaining the radio emission at early times in the fireball model
with a wind scaling for the external medium:
the peak frequency $\nu_{\rm m}$ (which corresponds to the minimum 
energy of the electron energy distribution), and the synchrotron 
self-absorption frequency $\nu_{\rm sa}$.
The case where $\nu_{\rm sa} < \nu_{\rm m} < \nu_{\rm c}$ can be
used to explain the radio evolution of SN~1978K.
In a snapshot of the spectrum, the flux rises as $\nu^2$ up to 
$\nu_{\rm sa}$, breaks to $\nu^{1/3}$ up to $\nu_{\rm m}$, falls 
as $\nu^{-0.687}$ up to $\nu_{\rm c}$, and falls as $\nu^{-1.187}$ 
beyond $\nu_{\rm c}$.
The four segments of the spectrum have different temporal evolutions:
$t^1$ up to $\nu_{\rm sa}$, $t^0$ between $\nu_{\rm sa}$ and 
$\nu_{\rm m}$, $t^{-1.53}$ between $\nu_{\rm m}$ and $\nu_{\rm c}$, 
and $t^{(2-3p)/4} = t^{-1.280}$ above $\nu_{\rm c}$.
The three characteristic frequencies are also not static, with
temporal evolutions
$\nu_{\rm sa} = A (t - t_0)^{-3/5}$,
$\nu_{\rm m} = B (t - t_0)^{-3/2}$,
and 
$\nu_{\rm c} = C (t - t_0)^{1/2}$.}

\edit1{From Figure 5, $\nu_{\rm c} = 54.55~{\rm GHz}$ at 
$(t - t_0) = 13934~{\rm days}$ giving $C = 0.4621$.
The flux density of 2.820~mJy at this time and this frequency fixes 
the overall normalization.}

\edit1{Figure \ref{figure:earlyradio}{(b)} shows the radio evolution 
of the fireball model at early times using $A = 88.8$ and 
$B = 3.727 \times 10^5$.
Unlike the \citet{Montes1997} model, the fireball model naturally
explains how the flux at 1.4~GHz remained constant when
$\nu_{\rm sa} < 1.4~{\rm GHz} < \nu_{\rm m}$, while at
the same time the flux at 0.8~GHz was rising as $t^1$ because
$0.8~{\rm GHz} < \nu_{\rm sa}$.
While this is promising independent evidence favoring the fireball
model, we caution that scintillation may be an issue at lower 
frequencies, there may be chromatic variability in the radio, 
and there may have been problems with the calibration or background 
subtraction in some of the early observations.}

\edit1{Figure \ref{figure:earlyradio}{(b)} shows the pure
unabsorbed radio emission from the fireball.
It appears to produce more than the observed flux at the lower
frequencies.
The same problem was found in the \citet{Montes1997} model, which
motivated including a time-independent and frequency-dependent
absorption from material that is ``distant'' from the source.
The same $e^{-\tau^{\prime\prime}}$ attenuation can be applied to
the fireball model, where 
$\tau^{\prime\prime} = K_4 (\nu / 5~{\rm GHz})^{-2.1}$ as in 
\citet{Montes1997}.
Figure \ref{figure:earlyradio}{(c)} shows an example of an
absorbed fireball with $A = 90.3$, $B = 3.287 \times 10^5$, 
and $K_4 = 7.0 \times 10^{-3}$: this appears to give a better
description of the early radio evolution, albeit with the same
caveats regarding the early radio observations at low frequencies.}

\edit1{For our recent ATCA and ALMA observations, the absorbed 
fireball model is shown as the dashed curves in Figures 
\ref{figure:atca} and \ref{figure:atcaalma}.
The 2013-06-07 ATCA data points lie $2.0\sigma$, $3.0\sigma$, 
$3.0\sigma$, and $3.4\sigma$ below the absorbed fireball model 
curves at 1.33~GHz, 1.84~GHz, 2.36~GHz, and 2.87~GHz respectively.
Thus these are much less discrepant than for the \citet{Montes1997} 
model, and there is less need to invoke chromatic variability
in the absorbed fireball model.}

\section{Dust Implications} \label{sec:dust}

Enormous amounts of dust (up to $10^{8} M_\odot$) have been found in 
high-redshift ($z \gtrsim 6$) galaxies from 
a variety of far-infrared and sub-millimeter observations 
\citep[e.g.][]{Pei1991,Pettini1997,Laporte2017}.
However, the source of this dust remains unclear, with several possible
contributors.

The short time scales required for dust enrichment make core-collapse
supernovae rather natural candidates for dust producers in the early 
Universe.
Estimates of the amount of dust produced per supernova
are sensitive to the choice of the initial mass function and the grain 
destruction efficiencies; it is likely that each supernova must 
produce 0.1--1~$M_\odot$ of dust to account for the high redshift 
observations \citep[e.g.][]{Dwek2007,Meikle2007}.  

Studies of young supernovae a few hundred days after their explosion
have tended to find dust masses that are much too low
\citep[e.g.][]{Kotak2009,Andrews2011,Meikle2011}, 
although the amount of dust may still be rising at this stage
\citep{Sarangi2013}.
On the other hand, studies of old supernova remnants have the problem
of the dust grains being destroyed by the reverse shock as the ejecta 
interacts with the ambient medium 
\citep[e.g.][]{Williams2006,Nozawa2006}.

Thus it is of particular interest to look for dust in supernovae in the
transitional phase that have not transitioned fully into the remnant phase. 
This ensures that ISM material has not been swept up by the ejecta, and 
that the emission is due to material directly associated with the 
progenitor system.  

\subsection{Warm Dust in SN~1978K} \label{subsec:warmdust}

The \Spitzer\ IRAC light curves in Figure \ref{figure:irac} show that
SN~1978K has faded steadily since the 2006 \AKARI\ observation.
This proves that this emission is coming from SN~1978K, and not from 
an unrelated dust cloud in the same direction.

\begin{figure}[ht!]
\plotone{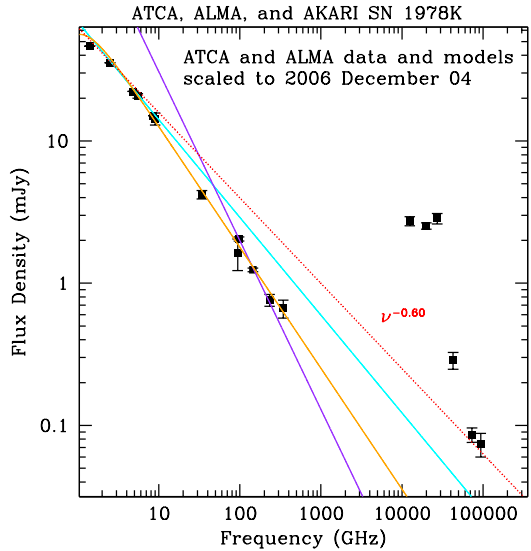}
\caption{The \AKARI\ infrared points show SN~1978K on 2006 December 03--04 
\citep{Tanaka2012}.
The model curves and ALMA and ATCA data points from 
Figure \ref{figure:atcaalma} have been extrapolated back to this time.
The dotted red $\nu^{-0.60}$ power law was previously used by
\citet{Tanaka2012} to explain their shortest wavelength observations.}
\label{figure:akarialmaatca}
\end{figure}

Figure \ref{figure:akarialmaatca} updates Figure 2 of \citet{Tanaka2012}.
The \AKARI\ infrared points show SN~1978K on 2006 December 03--04.
The model curves from Figure \ref{figure:atcaalma} have been extrapolated 
back to this time.
For the \citet{Montes1997} model, the evolution is $t^{-1.53}$ 
for all frequencies.
For the fireball model, the evolution is $t^{-1.53}$ below $\nu_{\rm c}$
and $t^{-1.28}$ above it.
The ALMA and ATCA data points from Figure \ref{figure:atcaalma} have been 
extrapolated back assuming the fireball evolution.

\citet{Tanaka2012} used a $\nu^{-0.60}$ power law -- shown as the 
dotted red curve in Figure \ref{figure:akarialmaatca} -- to
extrapolate from the radio to explain the shortest wavelength 
infrared emission at 3.2 and 4.1~\micron.
Instead, our ATCA and ALMA observations show that the 
radio-millimeter spectrum is much steeper than this.
Thus the contribution from synchrotron radiation to the infrared
emission is much smaller than assumed by \citet{Tanaka2012}.

\citet{Tanaka2012} preferred a model that used $1.3 \times 10^{-3}~ M_\sun$ 
of silicate dust at $T = 230~{\rm K}$.
This gives a double-peaked spectrum similar to the \Spitzer\ IRS one
shown in Figure \ref{figure:irs}.
Removing the synchrotron component, the increase in the 
peak flux that needs to come from warm dust is relatively small, 
$\lesssim 10$\%.
However, there is a much greater impact on the shape of the spectrum.
The favored silicate model shown in Figure 3 of \citet{Tanaka2012} 
is too narrow, with insufficient emission at 3.2 and 4.1~\micron.
Instead, the amorphous carbon model with $6.8 \times 10^{-3}~ M_\sun$
of dust at $T = 180~{\rm K}$ has a broader shape that could explain
both the 3.2 and 4.1~\micron\ emission and the longer wavelength emission,
but which fails to provide the peak at $\sim 10$~\micron.
This suggests that multiple dust emission components may be involved.

Figure \ref{figure:irac} shows that over the past decade, the 
fluxes at 3.2 and 4.1~\micron\ have dropped by a factor of $\sim 2$.
The $t^{-2.45}$ power law decay in both channels is much faster than
we have found in the radio ($t^{-1.53}$, Figure \ref{figure:atca}), 
optical, UV, or X-rays.
Since multiple dust emission components may be present, it should be 
noted that it is only the very warmest dust that is known to be fading.

The rapid decay at 3.2 and 4.1~\micron\ indicates this is not a simple 
geometric expansion of the emitting region.
Figure \ref{figure:irac} showed that the decay rate is the same 
in both infrared channels; this indicates that it is not the result 
of a cooling of the dust.
The simplest explanation is that dust is being destroyed.

\citet{Tanaka2012} preferred a forward shocked circumstellar dust 
model for the warm dust emission from SN~1978K, similar to SN~1987A.  
They claimed that an infrared echo model is unlikely, stating there is 
not enough energy for this.
However, they only considered an initial input of energy assuming the
supernova luminosity had an exponential decay with a characteristic
time of 25~days.
We have instead shown that SN~1978K has been a strong emitter of UV 
and X-rays over the past decades \citep{Smith2007a}.
The X-ray energization alone
($\sim 3 \times 10^{39}~\rm{ergs}~{\rm s}^{-1}$)
exceeds the infrared luminosity observed
($\sim 1.5 \times 10^{39}~\rm{ergs}~{\rm s}^{-1}$).
SN~1978K was first detected by \ROSAT\ 13 years after the explosion.
It has emitted a total of $\gtrsim 10^{48}$~ergs just in 
$0.2-10~{\rm keV}$ X-rays since then.
Given the dense medium around SN~1978K, an infrared echo model might 
still be relevant.
\edit1{However, it is still challenging to explain the different 
temporal decays in the infrared and X-rays.}

\subsection{Cold Dust in SN~1978K} \label{subsec:colddust}

Depending on the modeling of the synchrotron emission, it is possible 
that there is also a significant emission from cold dust in SN~1978K.


Figure \ref{figure:alma} shows that the Band 7 ALMA observations
can be fully explained using the \citet{Montes1997} model.
If this is the correct formalism then there is no indication of any
upturn in the spectrum at higher frequencies from a cold dust
component.

On the other hand, the fireball model seems to give a better 
description of the ALMA Band 3, 4, and 6 observations.
Figure \ref{figure:alma} shows that the Band 7 ALMA observations
lie a little above the $\nu^{-1.187}$ fireball curve.
Thus if the fireball model is the correct formalism then there is a 
suggestion that there may be an upturn in the spectrum at higher 
frequencies from a cold dust component.

\section{Conclusions and Future Work} \label{sec:conclusion}

We have shown that SN~1978K remains bright at longer wavelengths 40 
years after its explosion. 

Using updated ATCA light curves, we showed that, in general, the 
radio continues to decay as $t^{-1.53}$.

Our ALMA observations in 2016 show that the spectrum appears to be 
steeper at higher frequencies.
Two models can broadly explain all the radio and millimeter observations.
The \citet{Montes1997} model would require significant chromatic 
variability to explain the details.
Alternatively, an ultra-relativistic spherical blast wave in a wind 
scaling may be a better model.
Continued radio and millimeter monitoring could determine the correct 
model for the synchrotron spectrum by looking for the following
features:

\begin{itemize}

\item
For the \citet{Montes1997} model to be correct, it is necessary that 
the current ALMA Band 3 excess comes from a chromatic fluctuation.
Thus it is to be expected that this excess will recede in the next
few years.

\item
The fireball (or other broken power law) model predicts that the 
spectrum will always remain steeper in the millimeter than 
in the radio, with a distinct change in the spectral slope.
Observations in additional millimeter bands and/or with smaller
error bars in the individual SPW would better determine the
millimeter spectrum.

\item
The \citet{Montes1997} model predicts that the asymptotic temporal 
evolution of the spectrum is frequency independent ($t^{-1.53}$).
The fireball model instead predicts different temporal evolutions 
below $\nu_{\rm c}$ ($t^{-1.53}$) and above it ($t^{-1.28}$).

\item
If interstellar scintillation is causing significant variability in
the radio then the time scale for the variation should be $\sim $~hours 
to days \citep{Walker98}.
If long radio observations at multiple frequencies reveal different
variability at different frequencies, or sub-sets of the channels
within a band show different levels of variability, this would indicate
that scintillation is important.
On the other hand, if the radio variations are the result of 
inhomogeneities in the radio-emitting region, the time scale for 
the variations will be much longer and will depend on the mass-loss
evolution of the progenitor star.

\item
If scintillation is not a problem, then observations at radio 
frequencies below 1~GHz could determine the shape of the rollover 
in the spectrum and investigate the contribution of absorption 
from the distant surrounding medium.
The GaLactic and Extragalactic All-sky Murchison Widefield Array 
(GLEAM) survey \citep{Hurley-Walker2017} shows a source at the
location of SN~1978K with a peak brightness of 94~mJy/beam at 200~MHz.
This would be inconsistent with the rollover expected from absorption.
\edit1{At late times, the only significant absorption component 
remaining in the \citet{Montes1997} model comes from the $K_4$ term: 
this is the absorption far from the supernova that is supposed to be 
time independent, although in principle this term could be made more 
complex if the shock is assumed to have passed through part of 
the ``distant'' absorbing region.}
Currently, the large size of the GLEAM beam could include diffuse emission
from NGC~1313 and other confusing sources, so higher resolution
observations will be needed to cleanly separate SN~1978K from the 
host galaxy emission.

\end{itemize}

Supernovae such as SN~1978K are of particular interest since it has been 
suggested that they may be important contributors to the Universal 
dust budget. 
The strong X-ray and UV energization in SN~1978K suggests it
might not be a good place to create long-lasting dust.
However, SN~1978K does currently have a significant emission from 
warm dust: it was the only supernova in the transitional phase detected
by \citet{Tanaka2012}.

Our \Spitzer\ IRAC light curves show that at least
the warmest dust component has been decaying rapidly as $t^{-2.45}$ 
over the past decade.
This suggests this dust is currently being destroyed, although an 
infrared echo model for the warm dust emission cannot be discounted.
Monitoring the infrared emission -- and how it changes with the 
evolving X-ray and UV fluxes -- will be needed to distinguish
between the models for the dust emission.

We showed that there is a negligible contribution from synchrotron
radiation in the mid-infrared, and thus this emission must all
come from warm dust.
Currently, the amount of warm dust found in SN~1978K appears to be
$< 10^{-2}~ M_\sun$ \citep{Tanaka2012}.
Multiple dust emission components may be necessary, and 
better mid-infrared spectroscopy will be needed to determine the best 
model for the current composition, temperatures, and masses
of the warm dust.
This will potentially be possible using the {\it James Webb Space 
Telescope}.

Our ALMA observations have not so far revealed a bright emission from 
cold dust in SN~1978K.
However, this is more likely to be detected as an upturn in the spectrum
at higher sub-millimeter frequencies.
There is a large gap in frequencies between our 350~GHz ALMA observation
and the 12500~GHz \AKARI\ one that could potentially hide a significant
cold dust emission.
For example, the purple curves in Figure 10 of \citet{Tanaka2012} 
showed supernovae dust models with $T = 50~{\rm K}$;
scaling from these to the distance of SN~1978K, $1.0~M_\sun$ of cool
dust would produce a peak of $\sim 1$~mJy at $\sim 5000$~GHz.
Thus observations in ALMA Bands 8--10 are warranted when these modes 
have been fully standardized, to look for a significant upturn in the 
spectrum.

Detailed long-term observations of the decay of SN~1978K will also be 
important for comparing to other bright supernovae -- such as SN~1987A 
and SN~1996cr -- and will serve as a pathfinder for younger Type~IIn 
supernovae.

\acknowledgments

\edit1{We thank the referee for some important suggestions.}

ALMA is a partnership of ESO (representing its member states), NSF (USA) 
and NINS (Japan), together with NRC (Canada), MOST and ASIAA (Taiwan), 
and KASI (Republic of Korea), in cooperation with the Republic of Chile. 
The Joint ALMA Observatory is operated by ESO, AUI/NRAO and NAOJ.
The National Radio Astronomy Observatory is a facility of the National 
Science Foundation operated under cooperative agreement by Associated 
Universities, Inc.
This paper makes use of the following ALMA data: 
ADS/JAO.ALMA\#2015.1.00869.S (P.I. I. Smith).

The Australia Telescope Compact Array is part of the Australia Telescope 
National Facility which is funded by the Australian Government for 
operation as a National Facility managed by CSIRO.  
The ATCA data reported here were obtained under Program C184 
(P.I. S. Ryder).

This work is based in part on observations made with the {\it Spitzer 
Space Telescope}, which is operated by the Jet Propulsion Laboratory, 
California Institute of Technology under a contract with NASA.
The \Spitzer\ data reported here were obtained under Programs
10136, 11063, and 13053 (P.I. M. Kasliwal), 40204 (P.I. R. Kennicutt),
50603 (P.I. S. Van Dyk), 80015 (P.I. C. Kochanek),
\edit1{and 14108 (P.I. I. Smith).}

S.R. and R.K. acknowledge support from the Royal Society
International Exchange scheme (IE140343). 

%

\vspace{5mm}
\facilities{ALMA, ATCA, Spitzer, Akari.}


\software{CASA \citep{casa}, 
Miriad \citep{miriad},
MOPEX \citep{mopex}.}

\end{document}